\documentclass[12pt,a4paper]{article}

\usepackage[normalem]{ulem}
\usepackage{mathrsfs}
\usepackage{mathtools} 
\usepackage{multirow}
\usepackage{multicol}
\DeclarePairedDelimiter{\abs}{\lvert}{\rvert}  

\usepackage{ifthen} 
\newboolean{pdflatex}
\setboolean{pdflatex}{true} 

\newboolean{articletitles}
\setboolean{articletitles}{true} 

\newboolean{uprightparticles}
\setboolean{uprightparticles}{false} 

\newboolean{inbibliography}

\def\paperauthors{LHCb collaboration} 
\def\paperasciititle{Observation of CP violation in Bd2Jpsirho decays} 
\def\papertitle{Observation of $C\!P$ violation in $B^{0}\!\to{J\mskip-3mu/\mskip-2mu\psi}\rho(770)^0$ decays} 
\def\paperkeywords{{High Energy Physics}, {LHCb}} 
\def\papercopyright{\the\year\ CERN for the benefit of the LHCb collaboration} 
\def\paperlicence{CC BY 4.0 licence}
\def\paperlicenceurl{https://creativecommons.org/licenses/by/4.0/}

\newif\ifEnableSectionTOCLinks
\EnableSectionTOCLinksfalse 

\usepackage[top=1in, bottom=1.25in, left=1in, right=1in]{geometry}

\usepackage{microtype}
\usepackage{lineno}  
\usepackage{xspace} 
\usepackage{caption} 

\usepackage{graphicx}  
\usepackage{color}
\usepackage{colortbl}
\graphicspath{{./figs/}} 

\usepackage{amsmath} 
\usepackage{amssymb}
\usepackage{amsfonts}
\usepackage{upgreek} 

\newcommand*\patchAmsMathEnvironmentForLineno[1]{%
\expandafter\let\csname old#1\expandafter\endcsname\csname #1\endcsname
\expandafter\let\csname oldend#1\expandafter\endcsname\csname
end#1\endcsname
 \renewenvironment{#1}%
   {\linenomath\csname old#1\endcsname}%
   {\csname oldend#1\endcsname\endlinenomath}%
}
\newcommand*\patchBothAmsMathEnvironmentsForLineno[1]{%
  \patchAmsMathEnvironmentForLineno{#1}%
  \patchAmsMathEnvironmentForLineno{#1*}%
}
\AtBeginDocument{%
\patchBothAmsMathEnvironmentsForLineno{equation}%
\patchBothAmsMathEnvironmentsForLineno{align}%
\patchBothAmsMathEnvironmentsForLineno{flalign}%
\patchBothAmsMathEnvironmentsForLineno{alignat}%
\patchBothAmsMathEnvironmentsForLineno{gather}%
\patchBothAmsMathEnvironmentsForLineno{multline}%
\patchBothAmsMathEnvironmentsForLineno{eqnarray}%
}

\usepackage[pdftex,
            pdfauthor={\paperauthors},
            pdftitle={\paperasciititle},
            pdfkeywords={\paperkeywords}]{hyperref}
\usepackage{hyperxmp}
\hypersetup{
    pdfcopyright={Copyright (C) \papercopyright},
    pdflicenseurl={\paperlicenceurl}
}

\usepackage[colorinlistoftodos,textsize=scriptsize]{todonotes}

\usepackage[bottom,flushmargin,hang,multiple]{footmisc}

\usepackage[all]{hypcap} 

\usepackage{xspace} 
\usepackage{upgreek}

\def\lhcb   {\mbox{LHCb}\xspace}

\def\babar  {\mbox{BaBar}\xspace}
\def\belle  {\mbox{Belle}\xspace}

\def\MagUp {\mbox{\em Mag\kern -0.05em Up}\xspace}

\ifthenelse{\boolean{uprightparticles}}%
{

 \def\Peta        {\ensuremath{\upeta}\xspace}

 \def\Pmu         {\ensuremath{\upmu}\xspace}                 
 \def\Pnu         {\ensuremath{\upnu}\xspace}                 
                  
 \def\Ppi         {\ensuremath{\uppi}\xspace}                 
                  
 \def\Prho        {\ensuremath{\uprho}\xspace}

 \def\Ppsi        {\ensuremath{\uppsi}\xspace}

 \def\PDelta      {\ensuremath{\Delta}\xspace}                 
 \def\PXi         {\ensuremath{\Xi}\xspace}                 
 \def\PLambda     {\ensuremath{\Lambda}\xspace}                 
 \def\PSigma      {\ensuremath{\Sigma}\xspace}                 
 \def\POmega      {\ensuremath{\Omega}\xspace}                 
 \def\PUpsilon    {\ensuremath{\Upsilon}\xspace}
 \let\oldPi\Pi
 \def\PPi         {\ensuremath{\oldPi}\xspace}

 \def\PB      {\ensuremath{\mathrm{B}}\xspace}                 
 \def\PD      {\ensuremath{\mathrm{D}}\xspace}                 
 \def\PJ      {\ensuremath{\mathrm{J}}\xspace}                 
 \def\PK      {\ensuremath{\mathrm{K}}\xspace}                 
 \def\Pb      {\ensuremath{\mathrm{b}}\xspace}                 
 \def\Pc      {\ensuremath{\mathrm{c}}\xspace}                 
 \def\Pd      {\ensuremath{\mathrm{d}}\xspace}                 

 \def\Pp      {\ensuremath{\mathrm{p}}\xspace}                 
 \def\Pq      {\ensuremath{\mathrm{q}}\xspace}                 
 \def\Ps      {\ensuremath{\mathrm{s}}\xspace}

 \def\thebaroffset{0.0em}
}
{

 \def\Peta        {\ensuremath{\eta}\xspace}

 \def\Pmu         {\ensuremath{\mu}\xspace}                 
 \def\Pnu         {\ensuremath{\nu}\xspace}                 
                  
 \def\Ppi         {\ensuremath{\pi}\xspace}                 
                  
 \def\Prho        {\ensuremath{\rho}\xspace}

 \def\Ppsi        {\ensuremath{\psi}\xspace}                 
                  
 \mathchardef\PDelta="7101
 \mathchardef\PXi="7104
 \mathchardef\PLambda="7103
 \mathchardef\PSigma="7106
 \mathchardef\POmega="710A
 \mathchardef\PUpsilon="7107
 \mathchardef\PPi="7105
 \def\PB      {\ensuremath{B}\xspace}                 
 \def\PD      {\ensuremath{D}\xspace}                 
 \def\PJ      {\ensuremath{J}\xspace}                 
 \def\PK      {\ensuremath{K}\xspace}                 
 \def\Pb      {\ensuremath{b}\xspace}                 
 \def\Pc      {\ensuremath{c}\xspace}                 
 \def\Pd      {\ensuremath{d}\xspace}                 

 \def\Pp      {\ensuremath{p}\xspace}                 
 \def\Pq      {\ensuremath{q}\xspace}                 
 \def\Ps      {\ensuremath{s}\xspace}

 \def\thebaroffset{0.18em}
}
\newcommand{\offsetoverline}[2][\thebaroffset]{\kern #1\overline{\kern -#1 #2}}%

\makeatletter
\ifcase \@ptsize \relax
  \newcommand{\miniscule}{\@setfontsize\miniscule{4}{5}}
\or
  \newcommand{\miniscule}{\@setfontsize\miniscule{5}{6}}
\or
  \newcommand{\miniscule}{\@setfontsize\miniscule{5}{6}}
\fi
\makeatother

\DeclareRobustCommand{\optbar}[1]{\shortstack{{\miniscule (\rule[.5ex]{1.25em}{.18mm})}
  \\ [-.7ex] $#1$}}

\def\mup        {{\ensuremath{\Pmu^+}}\xspace}
\def\mun        {{\ensuremath{\Pmu^-}}\xspace} 

\def\mumu       {{\ensuremath{\Pmu^+\Pmu^-}}\xspace}

\def\quark     {{\ensuremath{\Pq}}\xspace}

\def\dquark    {{\ensuremath{\Pd}}\xspace}

\def\squark    {{\ensuremath{\Ps}}\xspace}

\def\cquark    {{\ensuremath{\Pc}}\xspace}
\def\cquarkbar {{\ensuremath{\overline \cquark}}\xspace}

\def\bquark    {{\ensuremath{\Pb}}\xspace}

\def\pion   {{\ensuremath{\Ppi}}\xspace}

\def\pip    {{\ensuremath{\pion^+}}\xspace}
\def\pim    {{\ensuremath{\pion^-}}\xspace}
\def\pipm   {{\ensuremath{\pion^\pm}}\xspace}

\def\rhomeson {{\ensuremath{\Prho}}\xspace}
\def\rhoz     {{\ensuremath{\rhomeson^0}}\xspace}

\def\kaon    {{\ensuremath{\PK}}\xspace}
\def\Kbar    {{\ensuremath{\offsetoverline{\PK}}}\xspace}

\def\KorKbar {\kern \thebaroffset\optbar{\kern -\thebaroffset \PK}{}\xspace}

\def\Kp      {{\ensuremath{\kaon^+}}\xspace}

\def\KS      {{\ensuremath{\kaon^0_{\mathrm{S}}}}\xspace}

\def\Kstarz  {{\ensuremath{\kaon^{*0}}}\xspace}
\def\Kstarzb {{\ensuremath{\Kbar{}^{*0}}}\xspace}

\newcommand{\etapr}{\ensuremath{\Peta^{\prime}}\xspace}

\def\D       {{\ensuremath{\PD}}\xspace}

\def\DorDbar {\kern \thebaroffset\optbar{\kern -\thebaroffset \PD}\xspace}

\def\Dp      {{\ensuremath{\D^+}}\xspace}
\def\Dm      {{\ensuremath{\D^-}}\xspace}

\def\DpDm    {\ensuremath{\Dp {\kern -0.16em \Dm}}\xspace}

\def\B       {{\ensuremath{\PB}}\xspace}
\def\Bbar    {{\ensuremath{\offsetoverline{\PB}}}\xspace}

\def\BorBbar {\kern \thebaroffset\optbar{\kern -\thebaroffset \PB}\xspace}

\def\Bd      {{\ensuremath{\B^0}}\xspace}
\def\Bdb     {{\ensuremath{\Bbar{}^0}}\xspace}
\def\BdorBdbar {\kern \thebaroffset\optbar{\kern -\thebaroffset \Bd}\xspace}
\def\Bu      {{\ensuremath{\B^+}}\xspace}

\def\Bs      {{\ensuremath{\B^0_\squark}}\xspace}
\def\Bsb     {{\ensuremath{\Bbar{}^0_\squark}}\xspace}
\def\BsorBsbar {\kern \thebaroffset\optbar{\kern -\thebaroffset \Bs}\xspace}
\def\Bc      {{\ensuremath{\B_\cquark^+}}\xspace}

\def\Bds     {{\ensuremath{\B_{(\squark)}^0}}\xspace}

\def\jpsi     {{\ensuremath{{\PJ\mskip -3mu/\mskip -2mu\Ppsi}}}\xspace}

\def\Y#1S{\ensuremath{\PUpsilon{(#1S)}}\xspace}

\def\proton      {{\ensuremath{\Pp}}\xspace}

\def\Lz          {{\ensuremath{\PLambda}}\xspace}

\def\LorLbar     {\kern \thebaroffset\optbar{\kern -\thebaroffset \PLambda}\xspace}

\def\Lb           {{\ensuremath{\Lz^0_\bquark}}\xspace}

\newcommand{\decay}[2]{\ensuremath{\mathinner{#1\!\to #2}}\xspace}

\def\to                 {\ensuremath{\rightarrow}\xspace}

\def\CP                {{\ensuremath{C\!P}}\xspace}

\newcommand{\dm}{{\ensuremath{\Delta m}}\xspace}
\newcommand{\dms}{{\ensuremath{\Delta m_{\squark}}}\xspace}
\newcommand{\dmd}{{\ensuremath{\Delta m_{\dquark}}}\xspace}
\newcommand{\DG}{{\ensuremath{\Delta\Gamma}}\xspace}
\newcommand{\DGs}{{\ensuremath{\Delta\Gamma_{\squark}}}\xspace}
\newcommand{\DGd}{{\ensuremath{\Delta\Gamma_{\dquark}}}\xspace}
\newcommand{\Gs}{{\ensuremath{\Gamma_{\squark}}}\xspace}

\newcommand{\DGq}{{\ensuremath{\Delta\Gamma_{\quark}}}\xspace}

\newcommand{\dmq}{{\ensuremath{\Delta m_{\quark}}}\xspace}

\newcommand{\phis}{{\ensuremath{\phi_{\squark}}}\xspace}

\def\BsToJPsiPhi  {\decay{\Bs}{\jpsi\phi}}

\def\AT#1     {\ensuremath{A_{\mathrm{T}}^{#1}}\xspace}           

\def\C#1      {\ensuremath{\mathcal{C}_{#1}}\xspace}                       
\def\Cp#1     {\ensuremath{\mathcal{C}_{#1}^{'}}\xspace}                    
\def\Ceff#1   {\ensuremath{\mathcal{C}_{#1}^{\mathrm{(eff)}}}\xspace}        
\def\Cpeff#1  {\ensuremath{\mathcal{C}_{#1}^{'\mathrm{(eff)}}}\xspace}       
\def\Ope#1    {\ensuremath{\mathcal{O}_{#1}}\xspace}                       
\def\Opep#1   {\ensuremath{\mathcal{O}_{#1}^{'}}\xspace}                    

\newcommand{\mbracket}[1]{\ensuremath{\left(#1\right)}} 

\newcommand{\aunit}[1]{\ensuremath{\text{\,#1}}}       

\newcommand{\tev}{\aunit{Te\kern -0.1em V}\xspace}
\newcommand{\gev}{\aunit{Ge\kern -0.1em V}\xspace}
\newcommand{\mev}{\aunit{Me\kern -0.1em V}\xspace}
\newcommand{\kev}{\aunit{ke\kern -0.1em V}\xspace}
\newcommand{\ev}{\aunit{e\kern -0.1em V}\xspace}
 
\newcommand{\mevc}{\ensuremath{\aunit{Me\kern -0.1em V\!/}c}\xspace}
\newcommand{\gevc}{\ensuremath{\aunit{Ge\kern -0.1em V\!/}c}\xspace}
\newcommand{\mevcc}{\ensuremath{\aunit{Me\kern -0.1em V\!/}c^2}\xspace}
\newcommand{\gevcc}{\ensuremath{\aunit{Ge\kern -0.1em V\!/}c^2}\xspace}

\def\fb   {\ensuremath{\aunit{fb}}\xspace}
\def\invfb   {\ensuremath{\fb^{-1}}\xspace}

\def\deriv {\ensuremath{\mathrm{d}}}

\def\gsim{{~\raise.15em\hbox{$>$}\kern-.85em
          \lower.35em\hbox{$\sim$}~}\xspace}
\def\lsim{{~\raise.15em\hbox{$<$}\kern-.85em
          \lower.35em\hbox{$\sim$}~}\xspace}

\newcommand{\mean}[1]{\ensuremath{\left\langle #1 \right\rangle}} 

\def\sqs   {\ensuremath{\protect\sqrt{s}}\xspace}

\def\pt         {\ensuremath{p_{\mathrm{T}}}\xspace}

\def\degrees{\ensuremath{^{\circ}}\xspace}

\def\mrad{\aunit{mrad}\xspace}
\def\rad{\aunit{rad}\xspace}

\def\tell1  {TELL1\xspace}
\def\ukl1   {UKL1\xspace}

\newcommand{\lhcborcid}[1]{\href{https://orcid.org/#1}{\hspace*{0.1em}\raisebox{-0.45ex}{\includegraphics[width=1em]{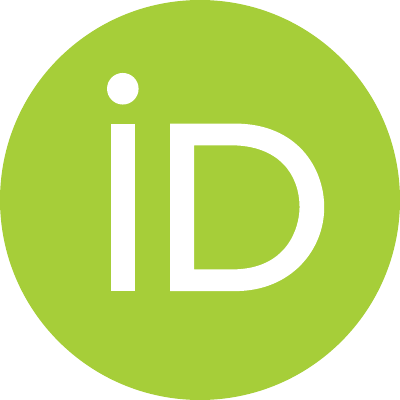}}}}

\hypersetup{
  colorlinks   = true, 
  urlcolor     = blue, 
  linkcolor    = blue, 
  citecolor    = red   
}

\ifEnableSectionTOCLinks
    \usepackage[explicit]{titlesec} 
    
    \let\oldcontentsline\contentsline
    \renewcommand

    \titleformat{\section}{\normalfont\Large\bf}{\hyperlink{tocsection.\thesection}{{\thesection} \parbox[t]{\dimexpr\textwidth-1pc}{#1}}}{1pc}{}

    \titleformat{\subsection}{\normalfont\bf}{\hyperlink{tocsubsection.\thesubsection}{{\thesubsection} \parbox[t]{\dimexpr\textwidth-1pc}{#1}}}{1pc}{}

    \titleformat{name=\section,numberless}[display]{}{}{0pt}{\normalfont\Huge\bfseries #1}
\fi

\usepackage{cite} 
\usepackage{mciteplus}
\mciteErrorOnUnknownfalse

\usepackage{amsmath}

\begin{document}

\renewcommand{\thefootnote}{\fnsymbol{footnote}}
\setcounter{footnote}{1}


\begin{titlepage}
\pagenumbering{roman}

\vspace*{-1.5cm}
\centerline{\large EUROPEAN ORGANIZATION FOR NUCLEAR RESEARCH (CERN)}
\vspace*{1.5cm}
\noindent
\begin{tabular*}{\linewidth}{lc@{\extracolsep{\fill}}r@{\extracolsep{0pt}}}
\ifthenelse{\boolean{pdflatex}}
{\vspace*{-1.5cm}\mbox{\!\!\!\includegraphics[width=.14\textwidth]{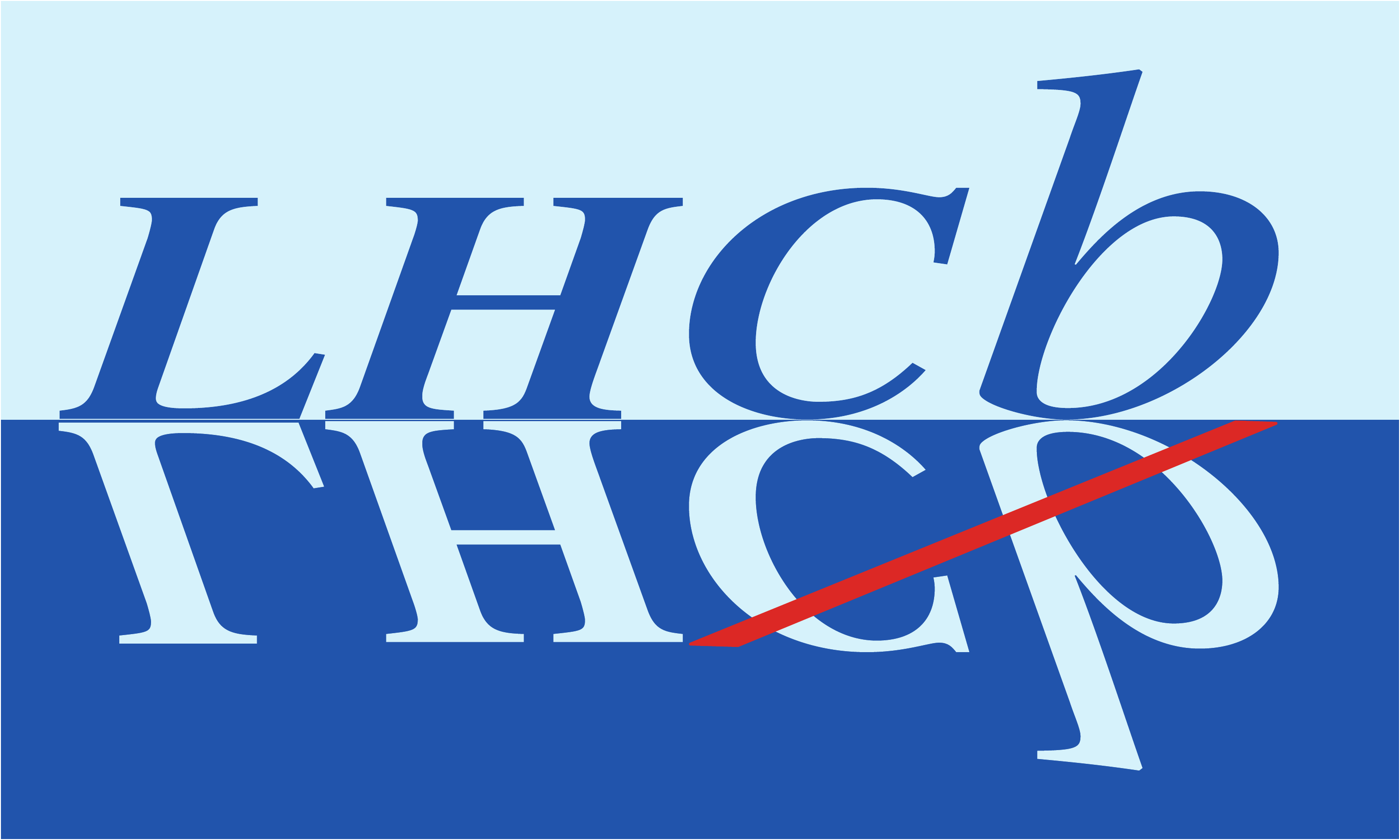}} & &}%
{\vspace*{-1.2cm}\mbox{\!\!\!\includegraphics[width=.12\textwidth]{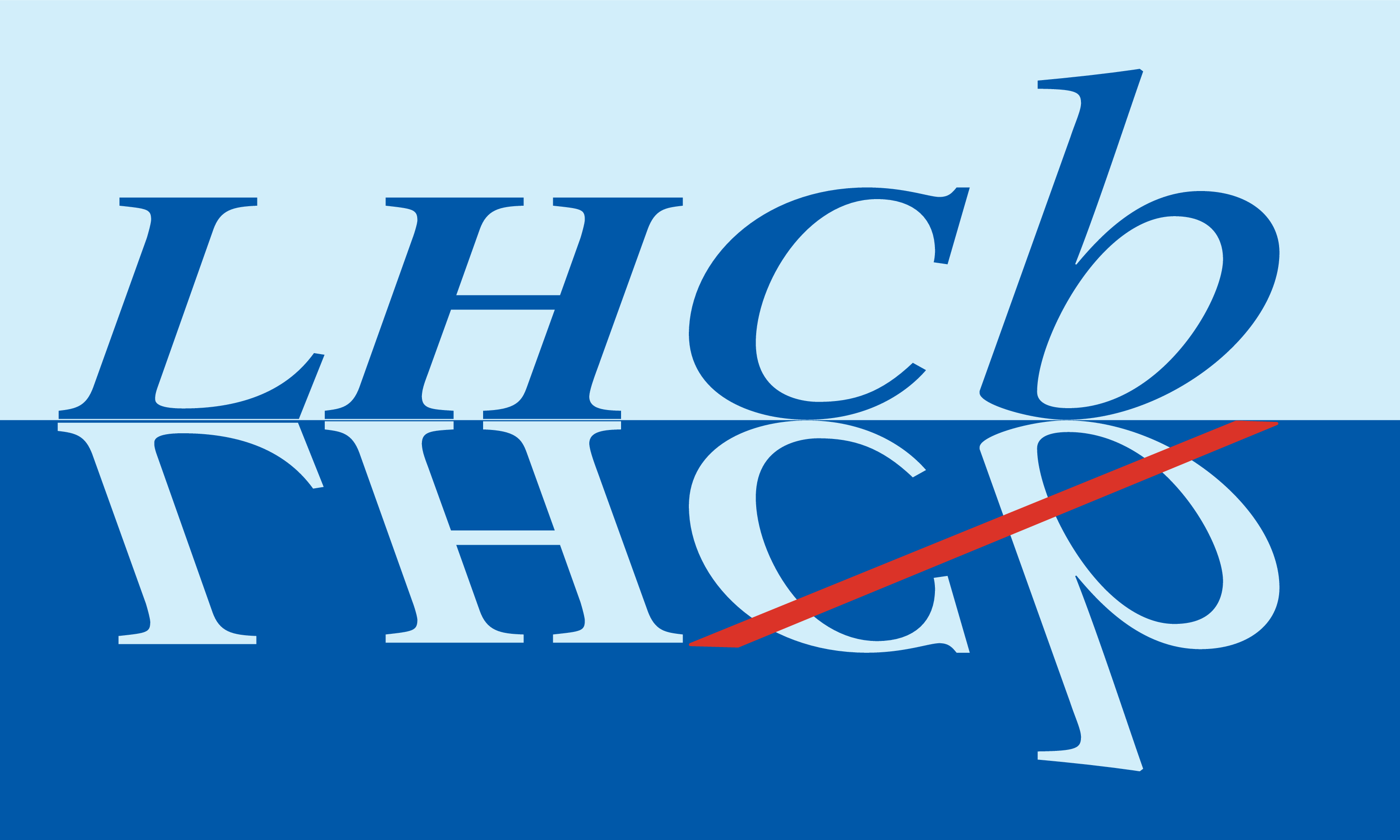}} & &}%
\\
 & & CERN-EP-2025-295 \\  
 & & LHCb-PAPER-2025-059 \\  
 & & July 16, 2026 \\ 
 & & \\
\end{tabular*}

\vspace*{2.0cm}

{\normalfont\bfseries\boldmath\huge
\begin{center}
  \papertitle 
\end{center}
}

\vspace*{2.0cm}

\begin{center}
\paperauthors\footnote{Authors are listed at the end of this paper.}
\end{center}

\vspace{\fill}

\begin{abstract}
  \noindent
   The time-dependent $C\!P$ asymmetry in $B^{0}\!\to{J\mskip-3mu/\mskip-2mu\psi}\rho(770)^0$ decays is measured using proton-proton collision data corresponding to an integrated luminosity of $6\,\text{fb}^{-1}$, collected with the LHCb detector at a center-of-mass energy of $13\,\text{Te\kern -0.1em V}$ during the years 2015--2018. The $C\!P$-violation parameters for this process are determined to be $2\beta^{\rm eff}_{c\bar{c}d} = 0.710 \pm 0.084 \pm 0.051\,\text{rad}$ and $|\lambda| = 1.019 \pm 0.034 \pm 0.024$, where the first uncertainty is statistical and the second systematic. This constitutes the first observation of time-dependent $C\!P$ violation in $B^{0}\!\to{J\mskip-3mu/\mskip-2mu\psi}\rho(770)^0$ decays. 
Assuming approximate SU(3) flavor symmetry, these results are combined with the previous consistent LHCb measurement to set the most stringent constraint on the penguin contribution, $\Delta\phi_{s}$, to the $C\!P$-violating phase $\phi_{s}$ in $B^{0}_{s}\!\to{J\mskip-3mu/\mskip-2mu\psi}\phi(1020)$ decays, yielding $\Delta\phi_{s} = 5.0 \pm 4.6\,\text{mrad}$. 
\end{abstract}

\vspace*{1.0cm}

\begin{center}
  Published in Phys.~Rev.~Lett. 137 (2026) 031803 
\end{center}

\vspace{\fill}

{\footnotesize 
\centerline{\copyright~\papercopyright. \href{\paperlicenceurl}{\paperlicence}.}}
\vspace*{2mm}

\end{titlepage}


\newpage
\setcounter{page}{2}
\mbox{~}


\renewcommand{\thefootnote}{\arabic{footnote}}
\setcounter{footnote}{0}


\cleardoublepage


\pagestyle{plain} 
\setcounter{page}{1}
\pagenumbering{arabic}


\def\les    {$<$\,}
\def\gre    {$>$\,}
\def\bel    {$\in$\,}

\def\thek    {\ensuremath{\mathrm{\theta}_{K}}\xspace}
\def\themu   {\ensuremath{\mathrm{\theta_{\mu}}}\xspace}
\def\thepi   {\ensuremath{\mathrm{\theta_{\pi}}}\xspace}
\def\theh    {\ensuremath{\mathrm{\theta}_{h}}\xspace}
\def\thel    {\ensuremath{\mathrm{\theta}_{l}}\xspace}
\def\phih    {\ensuremath{\mathrm{\phi}_{h}}\xspace}
\def\cosk    {\ensuremath{\mathrm{\cos\thek}}\xspace}
\def\cosmu   {\ensuremath{\mathrm{\cos\themu}}\xspace}
\def\cospi   {\ensuremath{\mathrm{\cos\thepi}}\xspace}
\def\coshd   {\ensuremath{\mathrm{\cos\theh}}\xspace}
\def\cosl    {\ensuremath{\mathrm{\cos\thel}}\xspace}

\newcommand{\tprime}{{\ensuremath{t^{\prime}}}\xspace}

\newcommand{\Azero}{{\ensuremath{A_{0}}}\xspace}
\newcommand{\Apar}{{\ensuremath{A_{\parallel}}}\xspace}
\newcommand{\Aperp}{{\ensuremath{A_{\perp}}}\xspace}
\newcommand{\Azerostar}{{\ensuremath{A_{0}^{*}}}\xspace}
\newcommand{\Aparstar}{{\ensuremath{A_{\parallel}^{*}}}\xspace}
\newcommand{\Aperpstar}{{\ensuremath{A_{\perp}^{*}}}\xspace}

\newcommand{\AS}{{\ensuremath{A_{S}}}\xspace}
\newcommand{\delzero}{{\ensuremath{\delta_{0}}}\xspace}
\newcommand{\delpar}{{\ensuremath{\delta_{\parallel}}}\xspace}
\newcommand{\delperp}{{\ensuremath{\delta_{\perp}}}\xspace}
\newcommand{\delS}{{\ensuremath{\delta_{S}}}\xspace}

\newcommand{\phizero}{{\ensuremath{\phi_{0}}}\xspace}
\newcommand{\phipar}{{\ensuremath{\phi_{\parallel}}}\xspace}
\newcommand{\phiperp}{{\ensuremath{\phi_{\perp}}}\xspace}
\newcommand{\phiS}{{\ensuremath{\phi_{S}}}\xspace}

\newcommand{\lamzero}{{\ensuremath{\lambda_{0}}}\xspace}
\newcommand{\lampar}{{\ensuremath{\lambda_{\parallel}}}\xspace}
\newcommand{\lamperp}{{\ensuremath{\lambda_{\perp}}}\xspace}
\newcommand{\lamS}{{\ensuremath{\lambda_{S}}}\xspace}

\newcommand{\pdfp}{{\ensuremath{\mathcal{P}}}\xspace}

\newcommand{\veca}{{\ensuremath{\vec{a}}}\xspace}
\newcommand{\vecx}{{\ensuremath{\vec{x}}}\xspace}
\newcommand{\vecz}{{\ensuremath{t}}\xspace}
\newcommand{\vecy}{{\ensuremath{\vec{\Omega}}}\xspace}
\newcommand{\vecT}{{\ensuremath{\vec{T}}}\xspace}
\newcommand{\csp}{{\ensuremath{C_{SP}}}\xspace}
\newcommand{\cspi}{{\ensuremath{C_{SP}^{i}}}\xspace}


\newcommand{\Af}{\ensuremath{\mathcal{A}_f}\xspace}
\newcommand{\Ak}{\ensuremath{\mathcal{A}_k}\xspace}
\newcommand{\Abarf}{\ensuremath{\bar{\mathcal{A}}_f}\xspace}
\newcommand{\Abark}{\ensuremath{\bar{\mathcal{A}}_k}\xspace}
\newcommand{\Afbar}{\ensuremath{\mathcal{A}_{\bar{f}}}\xspace}
\newcommand{\Akbar}{\ensuremath{\mathcal{A}_{\bar{k}}}\xspace}
\newcommand{\Abarfbar}{\ensuremath{\bar{\mathcal{A}}_{\bar{f}}}\xspace}
\newcommand{\Abarkbar}{\ensuremath{\bar{\mathcal{A}}_{\bar{k}}}\xspace}

\newcommand{\Afa}{\ensuremath{A_f}\xspace}
\newcommand{\Afb}{\ensuremath{\bar{A}_f}\xspace}
\newcommand{\Afbara}{\ensuremath{A_{\bar{f}}}\xspace}
\newcommand{\Afbarb}{\ensuremath{\bar{A}_{\bar{f}}}\xspace}


\newcommand{\invgev}{\ensuremath{\gev^{-1}}\xspace}
\newcommand{\gevt}{\ensuremath{\mathrm{Ge\kern -0.1em V}}\xspace}
\newcommand{\invgevc}{\ensuremath{{(\gevt\!/c)}^{-1}}\xspace}
\newcommand{\grad}{\ensuremath{^{\circ}}}

\def\BuToJPsiK {\decay{\Bu}{\jpsi\Kp}\xspace}

\def\BsJphi {\BsToJPsiPhi}
\def\Jmm {\jpsi(\mumu)}
\def\BsJPsimmPhi {\decay{\Bs}{\jpsi(\mumu)\phi(KK)}\xspace}
\def\BcJPsimunu {\decay{\Bc}{\Jmm \mup \Pnu}}
\newcommand{\CL}{C.L.\ }
\newcommand{\CLsb}{\ensuremath{\textrm{CL}_{\textrm{s+b}}}\xspace}
\newcommand{\CLs}{\ensuremath{\textrm{CL}_{\textrm{s}}}\xspace}
\newcommand{\CLb}{\ensuremath{\textrm{CL}_{\textrm{b}}}\xspace}
\newcommand{\MC}{Monte Carlo\xspace}

\newcommand{\mkk}{\ensuremath{m_{KK}}\xspace}
\newcommand{\mpipi}{\ensuremath{m_{\pi\pi}}\xspace}
\newcommand{\mhh}{\ensuremath{m_{hh}}\xspace}
\newcommand{\bbdim}{\ensuremath{b\bar{b}\to \mu \mu X}\xspace}
\newcommand{\mbdim}{\ensuremath{pp\to \mu \mu X}\xspace}

\newcommand{\Lambdab}{\ensuremath{\Lambda^0_b}\xspace}
\newcommand{\Bsmumu}{\ensuremath{\Bs\to\mu^+\mu^-}\xspace}
\newcommand{\Jpsimm}{\ensuremath{\Jpsi\to\mu^+\mu^-}\xspace}
\newcommand{\Bdmumu}{\ensuremath{\Bd\to\mu^+\mu^-}\xspace}
\newcommand{\KstKpi}{\ensuremath{\Kst\to K^+\pi^-}\xspace}
\newcommand{\DKpi}{\ensuremath{\D\to K^-\pi^+}\xspace}
\newcommand{\Bsmumugamma}{\ensuremath{\Bs\to\mu^+\mu^-\gamma}\xspace}
\newcommand{\Bdpipi}{\ensuremath{\Bd\to\pi^+\pi^-}\xspace}
\newcommand{\Bspipi}{\ensuremath{\Bs\to\pi^+\pi^-}\xspace}
\newcommand{\BsKK}{\ensuremath{\Bs\to K^+K^-}\xspace}
\newcommand{\BdKK}{\ensuremath{\Bd\to K^+K^-}\xspace}
\newcommand{\BdKpi}{\ensuremath{\Bd\to K^+\pi^-}\xspace}
\newcommand{\BspiK}{\ensuremath{\Bs\to\pi^+K^-}\xspace}
\newcommand{\Bhh}{\ensuremath{B^0_{(s)}\to h^+h^-}\xspace}
\newcommand{\Bmm}{\ensuremath{B^0_{(s)}\to \mu^+\mu^-}\xspace}
\newcommand{\Buhhh}{\ensuremath{B^{\pm}\to h^+h^-h^{\pm}}\xspace}
\newcommand{\BsDspi}{\ensuremath{B^0_{(s)}\to  D^-_s(K^+K^-\pi^-) \pi^+}\xspace}\newcommand{\Kst}{\ensuremath{K^{*0}}\xspace}
\newcommand{\BJpsiX}{\ensuremath{B\to J/\psi X}\xspace}
\newcommand{\BuJpsiK}{\ensuremath{B^+\to J/\psi K^+}\xspace}
\newcommand{\BuJpsimumuK}{\ensuremath{B^+\to J/\psi(\mu^+\mu^-)K^+}\xspace}
\newcommand{\BJpsimumuX}{\ensuremath{B \to J/\psi(\mu^+\mu^-)X}\xspace}
\newcommand{\BdJpsiKst}{\ensuremath{B^0_d\to J/\psi K^{*0}}\xspace}
\newcommand{\BdJpsiKpi}{\ensuremath{B^0_d\to J/\psi K \pi}\xspace}
\newcommand{\BJpsiKpi}{\ensuremath{B^0_{d(s)}\to J/\psi K \pi}\xspace}
\newcommand{\BsJpsiKpi}{\ensuremath{B^0_s\to J/\psi K \pi}\xspace}
\newcommand{\BsJpsiKst}{\ensuremath{B^0_s\to J/\psi \Kstarzb}\xspace}
\newcommand{\BJpsiKst}{\ensuremath{B^0_{d(s)}\to J/\psi K^{*0}(\Kstarzb)}\xspace}
\newcommand{\BsKstKst}{\ensuremath{\Bs\to\Kstarz\Kstarzb}\xspace}
\newcommand{\BdJpsimumuKstKpi}{\ensuremath{B^0_d\to J/\psi(\mu^+\mu^-)K^{*0}(K^+\pi^-)}\xspace}
\newcommand{\BsJpsimumuKstKpi}{\ensuremath{B^0_s\to J/\psi(\mu^+\mu^-)K^{*0}(K^+\pi^-)}\xspace}
\newcommand{\BcJpsimumumunu}{\ensuremath{B^+_c\to J/\psi(\mu^+\mu^-)\mu^+\nu_{\mu}}\xspace}
\newcommand{\Jpsimumu}{\ensuremath{J/\psi\to \mu^+\mu^-}\xspace}
\newcommand{\Jpsi}{\ensuremath{J/\psi}\xspace}

\newcommand{\Bqmumu}{\ensuremath{\ensuremath{B^0_{q}}\to\mu^+\mu^-}\xspace}
\newcommand{\BsJpsimumuPhiKK}{\ensuremath{B^0_s\to J/\psi(\mu^+\mu^-) \phi(K^+K^-)}\xspace}
\newcommand{\BdJpsimumupipi}{\ensuremath{B^0_d\to J/\psi(\mu^+\mu^-) \pi^+\pi^-}\xspace}
\newcommand{\BdJpsiRho}{\ensuremath{B^0_d\to J/\psi \rho^0}\xspace}
\newcommand{\BdJpsipipi}{\ensuremath{B^0_d\to J/\psi \pi^+\pi^-}\xspace}
\newcommand{\BsJpsiKK}{\ensuremath{B^0_s\to J/\psi K^+K^-}\xspace}
\newcommand{\ropipi}{\ensuremath{\rho^0\to \pi^+\pi^-}\xspace}

\newcommand{\Lambdappi}{\ensuremath{\Lambda\to p\pi^-}\xspace}
\newcommand{\Lambdabppi}{\ensuremath{\Lambdab\to p\pi^-}\xspace}
\newcommand{\LambdabpK}{\ensuremath{\Lambdab\to p K^-}\xspace}
\newcommand{\LbJpsipK}{\ensuremath{\Lambdab\to\Jpsi p K^-}\xspace}

\newcommand{\BsJpsiPhi}{\ensuremath{B^0_s\to J/\psi \phi}\xspace}

\newcommand{\BRof}[1]{\ensuremath{{\cal B}(#1)}\xspace}

\newcommand{\MeVc}{\ensuremath{\,{\rm MeV}/c}\xspace}
\newcommand{\GeVc}{\ensuremath{\,{\rm GeV}/c}\xspace}
\newcommand{\MeVcc}{\ensuremath{\,{\rm MeV}/c^2}\xspace}
\newcommand{\GeVcc}{\ensuremath{\,{\rm GeV}/c^2}\xspace}
\newcommand{\microb}{\ensuremath{\,{\rm \mu b}}\xspace}

\newcommand{\Enow}{\ensuremath{\,{\sqrt{s}=7\TeV}}\xspace}
\newcommand{\Einj}{\ensuremath{\,{\sqrt{s}=900\GeV}}\xspace}
\newcommand{\Enom}{\ensuremath{\,{\sqrt{s}=14\TeV}}\xspace}
\newcommand{\Efive}{\ensuremath{\,{\sqrt{s}=10\TeV}}\xspace}
\newcommand{\pdf}{\ensuremath{p.d.f.}\xspace}
\newcommand{\pdfs}{\ensuremath{p.d.f.s}\xspace}
\newcommand{\IP}{\ensuremath{{\rm IP}}\xspace}
\newcommand{\ThetaK}{\ensuremath{{\Theta_{K^{*}}}}\xspace}
\newcommand{\fdfs}{\ensuremath{\frac{f_d}{f_s}}\xspace}

\newcommand{\DLL}{\ensuremath{{\rm \Delta LL}}\xspace}
\newcommand{\gl}{\ensuremath{{\rm GL}}\xspace}
\newcommand{\glk}{\ensuremath{{\rm GL_K}}\xspace}
\newcommand{\glksm}{\ensuremath{{\rm GL_{KS}}}\xspace}
\newcommand{\glsm}{\ensuremath{{\rm GL_{S}}}\xspace}
\newcommand{\PID}{\ensuremath{{\rm PID}}\xspace}
\newcommand{\etos}{\ensuremath{\epsilon^{TOS/SEL}}\xspace}
\newcommand{\etis}{\ensuremath{\epsilon^{TIS/SEL}}\xspace}
\newcommand{\etistos}{\ensuremath{\epsilon^{TIS\&TOS/SEL}}\xspace}

\newcommand{\etrig}{\ensuremath{\epsilon^{TRIG/SEL}}\xspace}
\newcommand{\eSelect}{\ensuremath{\epsilon^{SEL/REC}}\xspace}
\newcommand{\eReco}{\ensuremath{\epsilon^{REC}}\xspace}

\newcommand{\sumpt}{\ensuremath{{\Sigma \pt}}\xspace}
\newcommand{\maxpt}{\ensuremath{{\rm max_{\pt}}}\xspace}
\newcommand{\maxip}{\ensuremath{\rm max_{\IP}}\xspace}

\newcommand{\swave}{{\em S--wave}\xspace}
\newcommand{\pwave}{{\em P--wave}\xspace}
\newcommand{\dwave}{{\em D--wave}\xspace}
\newcommand{\hwave}{{\em H--wave}\xspace}
\newcommand{\fwave}{{\em F--wave}\xspace}

\newcommand\Tstrut{\rule{0pt}{2.6ex}}
\newcommand\TTstrut{\rule{0pt}{3.2ex}}
\newcommand\Bstrut{\rule[-1.2ex]{0pt}{0pt}}
\newcommand\BBstrut{\rule[-1.8ex]{0pt}{0pt}}
\newcommand{\rhoJ}{\ensuremath{\rho_{\Jpsi}}\xspace}
\newcommand{\figref}[1]{Fig.~\ref{#1}}
\newcommand{\tabref}[1]{Table~\ref{#1}}
\newcommand{\appref}[1]{Appendix~\ref{#1}}
\newcommand{\bibref}[1]{Ref.~\cite{#1}}
\newcommand{\secref}[1]{Sect.~\ref{#1}}

\newcommand{\tred}[1]{\textcolor{red}{#1}}
\newcommand{\phisi}[1][i]{\phi_\text{s}^{#1}}
\newcommand{\phisav}{\phisi[\text{av}]}
\newcommand{\Delphisi}[1][i]{\Delta\phis^{#1}}
\newcommand{\Delphispara}{\Delphisi[\parallel]}
\newcommand{\Delphisperp}{\Delphisi[\perp]}
\newcommand{\Delphisperpp}{\Delphisi[\perp\prime]}
\newcommand{\DelphisS}{\Delphisi[\text{S}]}
\newcommand{\lamsAbs}{\lamfAbs[\text{s}]}
\newcommand{\lamsi}[1][i]{{\lamf[\text{s}]^{#1}}}
\newcommand{\lamsiAbs}[1][i]{|{\lamsi[#1]}|}
\newcommand{\lamsiSq}[1][i]{{\lamsiAbs[#1]}^2}
\newcommand{\Cs}{C_\text{s}}
\newcommand{\Csi}[1][i]{C_\text{s}^{#1}}
\newcommand{\Cszero}{\Csi[\text{0}]}
\newcommand{\Cspar}{\Csi[\parallel]}
\newcommand{\Csperp}{\Csi[\perp]}
\newcommand{\CsS}{\Csi[\text{S}]}
\newcommand{\Csav}{\Cs^\text{av}}
\newcommand{\DelCspara}{\Delta\Cs^\parallel}
\newcommand{\DelCsperp}{\Delta\Cs^\perp}
\newcommand{\CsavS}{\Cs^\text{avS}}
\newcommand{\taus}{\tau_\text{s}}
\newcommand{\taud}{\tau_\text{d}}
\newcommand{\mL}{M_\text{L}}
\newcommand{\mH}{M_\text{H}}

\newcommand{\AAv}[1][\Ai]{#1^{\text{CP}}}
\newcommand{\AAvConj}[1][\Ai]{{\AAv[#1]}^\ast}
\newcommand{\eGst}{e^{-\Gs\,t}}
\newcommand{\eGstal}{e^{-\Gs\,\alert{t}}}
\newcommand{\cDG}{\cosh\!\left(\frac{\DG}{2}t\right)}
\newcommand{\sDG}{\sinh\!\left(\frac{\DG}{2}t\right)}
\newcommand{\cDm}{\cos\!\left(\dm\, t\right)}
\newcommand{\sDm}{\sin\!\left(\dm\, t\right)}
\newcommand{\cDGs}{\cosh\!\left(\frac{\DGs}{2}t\right)}
\newcommand{\sDGs}{\sinh\!\left(\frac{\DGs}{2}t\right)}
\newcommand{\cDms}{\cos\!\left(\dms\, t\right)}
\newcommand{\sDms}{\sin\!\left(\dms\, t\right)}
\newcommand{\sDGd}{\sinh\!\left(\frac{\DGd}{2}t\right)}
\newcommand{\cDGd}{\cosh\!\left(\frac{\DGd}{2}t\right)}
\newcommand{\cDGq}{\cosh\!\left(\frac{\DGq}{2}t\right)}
\newcommand{\cDmd}{\cos\!\left(\dmd\, t\right)}
\newcommand{\cDmq}{\cos\!\left(\dmq\, t\right)}
\newcommand{\sDmd}{\sin\!\left(\dmd\, t\right)}
\newcommand{\scDms}{\cos}
\newcommand{\ssDms}{\sin}
\newcommand{\magzero}{|\Azero|}
\newcommand{\magzeroSq}{\magzero^2}
\newcommand{\magzeroAv}{|\AAv[\Azero]|}
\newcommand{\magzeroAvSq}{\magzeroAv^2}
\newcommand{\magpar}{|\Apar|}
\newcommand{\magparSq}{\magpar^2}
\newcommand{\magparAv}{|\AAv[\Apar]|}
\newcommand{\magparAvSq}{\magparAv^2}
\newcommand{\magperp}{|\Aperp|}
\newcommand{\magperpSq}{\magperp^2}
\newcommand{\magperpAv}{|\AAv[\Aperp]|}
\newcommand{\magperpAvSq}{\magperpAv^2}
\newcommand{\magS}[1][]{|A_{\text{S}#1}|}
\newcommand{\magSSq}[1][]{\magS[#1]^2}
\newcommand{\magSAv}[1][]{|\AAv[\AS]|}
\newcommand{\magSAvSq}[1][]{\magSAv^2}
\newcommand{\FS}[1][]{F_{\text{S}#1}}
\newcommand{\FSAv}[1][]{\FS[#1]^\text{CP}}

\newcommand{\delSzero}[1][]{\delS[#1]\,\text{--}\,\delzero[#1]}
\newcommand{\delSperp}[1][]{\delS[#1]\,\text{--}\,\delperp[#1]}
\newcommand{\delSpar}[1][]{\delS[#1]\,\text{--}\,\delpar[#1]}
\newcommand*\rot{\rotatebox{270}}
\newcommand{\tabcaption}[1]{  
\vspace{-\abovecaptionskip} %
\caption[#1]{#1}
\vspace{\abovecaptionskip}
}

\newcommand{\pp}{{\ensuremath{p{\xspace}p}}\xspace}
\newcommand{\lamf}{\ensuremath{\lambda_{f}}\xspace}
\newcommand{\lamfbar}{\ensuremath{\lambda_{\bar{f}}}\xspace}
\newcommand{\lamfp}{\ensuremath{\lambda_{f}^{\prime}}\xspace}
\newcommand{\lam}{\ensuremath{\lambda}\xspace}
\newcommand{\lamp}{\ensuremath{\xi}\xspace}
\newcommand{\betaeff}{{\ensuremath{\beta^{\rm eff}_{c\bar{c}d}}}\xspace}
\newcommand{\betaccs}{{\ensuremath{\beta^{\rm eff}_{c\bar{c}s}}}\xspace}
\newcommand{\betaeffo}{{\ensuremath{\beta^{\rm eff}_{\rm other}}}\xspace}
\newcommand{\betaeffz}{{\ensuremath{\beta^{\rm eff}_{0}}}\xspace}
\newcommand{\betaeffl}{{\ensuremath{\beta^{\rm eff}_{\parallel}}}\xspace}
\newcommand{\betaeffe}{{\ensuremath{\beta^{\rm eff}_{\perp}}}\xspace}
\newcommand{\betaeffi}{{\ensuremath{\beta^{\rm eff}_{i}}}\xspace}

\newcommand{\IndStaRho}{0.203}
\newcommand{\ValDphis}{5.0\pm4.6}

Studies of $\CP$ violation in $B$ mesons are essential for understanding the matter--antimatter asymmetry in the Universe~\cite{Sakharov:1967dj,PDG2024}. This phenomenon was first observed in $\Bd$ meson by the \babar~\cite{BaBar:2001pki} and \belle~\cite{Belle:2001zzw} experiments. Such measurements play a vital role in testing the Standard Model (SM) and exploring new physics phenomena.
The \CP-violating phase associated with the $\Bs$--$\Bsb$ mixing process, $\phis$, is a key observable sensitive to contributions of new particles in the mixing~\cite{Buras:2009if}. In the SM, its value at tree level is precisely determined by a global fit to experimental data to be ${-37.6}^{+0.6}_{-0.5}\mrad$~\cite{CKMfitter2015}.
It is measured through the interference between $\Bs$--$\Bsb$ mixing and subsequent decays 
via $\decay{\bquark}{\cquark\cquarkbar\squark}$ transitions, neglecting subleading electroweak-loop (penguin) contributions. Current measurements in the golden decay channel 
$\decay{\Bs}{\jpsi\phi(1020)}$ yield an average value of \mbox{$\phis=-50\pm17\mrad$}~\cite{LHCb-PAPER-2023-016,ATLAS:2020lbz,CMS:2024znt}, 
with an experimental uncertainty now comparable to the possible shift, $\Delta\phis$, due to the neglected penguin contamination with a different weak phase from the leading tree amplitude~\cite{Liu:2013nea,DeBruyn:2014oga,Frings:2015eva,Barel:2020jvf,DeBruyn:2025rhk}.
Thus, a thorough understanding of the penguin effects becomes mandatory for precision tests of the SM.

Long-distance nonperturbative effects in $\decay{\Bs}{\jpsi\phi(1020)}$ decays prevent a precise 
theoretical calculation of $\Delta\phis$. 
An estimate of $\Delta\phis$ can be obtained using \CP-violation parameters in $\decay{\bquark}{\cquark\cquarkbar\squark}$ and $\decay{\bquark}{\cquark\cquarkbar\dquark}$ decays employing SU(3) flavor symmetry, assuming that the corresponding ratios of the hadronic amplitudes of the penguin and tree decay topologies are 
identical~\cite{Liu:2013nea,DeBruyn:2014oga,Frings:2015eva,Barel:2020jvf,DeBruyn:2025rhk}.  
The decay $\decay{\Bd}{\jpsi\rho(770)^0}$, which proceeds via the Cabibbo-suppressed $\decay{\bquark}{\cquark\cquarkbar\dquark}$ transition, 
serves as a control channel due to its enhanced sensitivity to the hadronic parameters of penguin contributions relative to 
$\decay{\Bs}{\jpsi\phi(1020)}$~\cite{Barel:2020jvf}.

The measured \CP-violation parameters in $\decay{\Bd}{\jpsi\rho(770)^0}$ and $\decay{\Bd}{\jpsi\KS}$ decays allow for a determination of the penguin contributions in the former, which 
is used to estimate the penguin shift $\Delta\phis$. This estimate was performed by the LHCb collaboration 
using proton-proton ($\pp$) collision data collected in 2011--2012 \mbox{(Run~1)} at center-of-mass energies of $\sqs=7$ and $8\tev$,  corresponding to an integrated luminosity of \mbox{3\invfb}~\cite{LHCb-PAPER-2014-058}.
The obtained result of $\Delta\phis=0.9\pm9.8\mrad$ remains insufficiently precise to constrain the penguin effects on the $\phis$ measurement. 
This Letter reports an updated study of \CP violation in $\decay{\Bd}{\jpsi\rho(770)^0}$ decays, 
with $\decay{\jpsi}{\mup\mun}$ and $\decay{\rho(770)^0}{\pip\pim}$, using $\pp$ collision data collected by the \lhcb experiment during 2015--2018 \mbox{(Run~2)}, 
corresponding to an integrated luminosity of \mbox{6\invfb} at $\sqs=13\tev$, which leads to 
the first observation of \CP violation in this process. These results are combined with those of the \mbox{Run~1} measurement~\cite{LHCb-PAPER-2014-058} to determine $\Delta\phis$. 
As the analysis procedures for the $\decay{\Bd}{\jpsi\pip\pim}$ and $\decay{\Bs}{\jpsi\pip\pim}$ decays are similar, this study largely follows the procedures described in Refs.~\cite{LHCb-PAPER-2014-058,LHCb-PAPER-2019-003}.

The \lhcb detector is a single-arm forward spectrometer covering the \mbox{pseudorapidity} range $2<\eta <5$, 
described in detail in Refs.~\cite{LHCb-DP-2008-001,LHCb-DP-2014-002}. Simulated events are generated with 
the software described in Refs.~\cite{Sjostrand:2007gs,*Sjostrand:2006za,LHCb-PROC-2010-056,Lange:2001uf,davidson2015photos}. 
They are tuned to match data-taking conditions and are used to study selection requirements, as well as to model detector acceptance and resolution effect.
The trigger system performs the online selection~\cite{LHCb-DP-2012-004} via a hardware stage  
with requirements of high transverse momentum of hadrons, followed by a software stage 
using two trigger algorithms based on the muon kinematics and two-muon vertex information. 
Since only the first algorithm requires a large impact parameter significance relative to $\pp$ interaction point (primary vertex), 
the two selections exhibit different efficiency dependencies on the $\Bd$ decay time, decay angles, and the $\pip\pim$ invariant mass. Consequently, these dependencies are studied separately. 

The reconstruction and selection of $\decay{\Bd}{\jpsi(\mup\mun)\pip\pim}$ candidates, which include contributions from the $\rho(770)^0$ and other resonances in the full $\pip\pim$ invariant-mass spectrum, follow the strategy of Ref.~\cite{LHCb-PAPER-2019-003}. 
A boosted decision tree classifier is employed to suppress combinatorial background, using variables with good discriminating power, in particular the fit quality of $\Bd$ decay vertex, as well as the particle-identification information and impact parameter of the pions~\cite{LHCb-DP-2008-001}. This selection improves the signal purity under the $\Bd$ peak, defined as within $\pm20\mevcc$ of the known \Bd mass~\cite{PDG2024}, by approximately a factor of eight. 
Peaking backgrounds due to kaon and proton misidentification from 
$\decay{\Bds}{\jpsi\Kst}$, $\decay{\Lb}{\jpsi\proton{h^{-}}(h=K,\pi)}$, and 
$\decay{\Bu}{\jpsi\Kp}$ decays are 
removed by stringent particle identification and mass requirements, while $\decay{\Bd}{\jpsi\KS}$ decays are vetoed using the $\pip\pim$ invariant mass. 
The same-sign decays $\decay{\Bd}{\jpsi\pi^\pm\pi^\pm}$, containing two pions with same charge and selected using the same criteria, are used to study the mass shape of the combinatorial background. 

Selected $\decay{\Bd}{\jpsi\pip\pim}$ candidates in the mass range $[5250, 5500]\mevcc$
are retained for further analysis. 
The invariant mass of a \Bd candidate, $m(\jpsi\pip\pim)$, is calculated by requiring the \Bd momentum vector 
to point back to the closest primary vertex and constraining the $\mup\mun$ invariant mass to the known \jpsi mass~\cite{PDG2024}.
An unbinned maximum-likelihood fit to the $m(\jpsi\pip\pim)$ distribution 
is performed to discriminate the signal from background contributions.  
The mass shape of the combinatorial background is described by a fifth-order polynomial function, 
which can account for a dip caused by the veto of the $\decay{\Bd}{\jpsi\Kst}$ decays. Its parameters are 
shared between the $\decay{\Bd}{\jpsi\pip\pim}$ and $\decay{\Bd}{\jpsi\pipm\pipm}$ data samples and 
obtained via a simultaneous fit to both samples. 

\begin{figure}[tb]
  \begin{center}
     \includegraphics[width=0.65\linewidth]{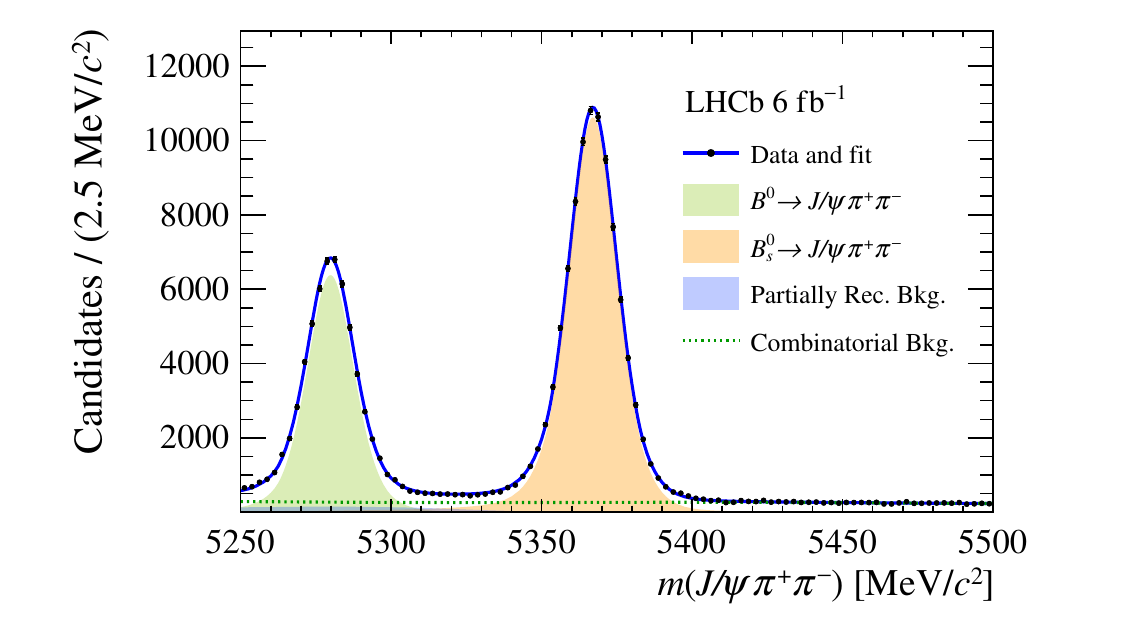}
     \vspace*{-0.5cm}
  \end{center}
  \caption{\small 
  Distribution of the $J\mskip -3mu/\mskip -2mu\psi\pi^{+}\pi^{-}$ invariant mass in the full \mbox{Run~2} data sample. Projections of the total fit and different contributions are also shown. 
  }
  \label{fig_jpsipipi}
\end{figure}

The \Bd and \Bs peaks are each modeled by the sum of a Hypatia function~\cite{Santos:2013gra} and
a Gaussian function, where the mass difference between \Bd and \Bs peaks is constrained to its known value~\cite{PDG2024}. 
Partially reconstructed backgrounds from $\decay{\Bs}{\jpsi\etapr}(\rhoz\gamma)$ and $\decay{\Bs}{\jpsi\eta}(\pip\pim\gamma)$ decays 
are included in the fit. Their shapes are modeled using simulations, and their yield ratios relative to the \Bd signal are constrained to the values determined from simulations. 
The resulting fit is shown in \figref{fig_jpsipipi} and is used to estimate the yield of each component under the \Bd peak.

A signal sample of approximately $51\,000$ $\decay{\Bd}{\jpsi\pip\pim}$ decays is obtained. In the subsequent analysis, the residual backgrounds under the $\Bd$ peak are statistically subtracted by injecting events with negative weights, with the total weight matching the corresponding background yield~\cite{LHCb-PAPER-2020-045,LHCb-PAPER-2019-003}. 
The combinatorial background is subtracted by using candidates in the mass 
interval $[5312,5332]\mevcc$, while simulated samples are used for the partially reconstructed backgrounds. The small contamination from the tail of the $\decay{\Bs}{\jpsi\pip\pim}$ decays
is subtracted using an amplitude-corrected simulation sample~\cite{LHCb-PAPER-2019-003}.

The measurement of \CP violation in $\decay{\Bd}{\jpsi\pip\pim}$ decays requires an amplitude analysis to separate different polarization components.  
A weighted multidimensional maximum-likelihood fit is performed to the background-subtracted distributions of the decay time, 
the invariant mass $m(\pip\pim)$, and the angular variables $(\cospi, \cosmu, \chi)$.  
The helicity angle \thepi (\themu) is defined as the angle between the \pip (\mup) momentum and the  
direction opposite to the $\Bd$ momentum in the $\pip\pim(\mup\mun)$ rest frame, 
and $\chi$ is the azimuthal angle between the $\pip\pim$ and $\mumu$ decay planes~\cite{LHCb-PAPER-2014-058,LHCb-PAPER-2020-042}.

The fit is performed simultaneously across six subsamples, categorized by data-taking period and trigger category, 
with a common set of physics parameters. 
The signal probability density function (PDF) in each subsample closely follows the formalism detailed in Refs.~\cite{LHCb-PAPER-2019-003,Zhang:2012zk}. 
It incorporates the effects of flavor tagging, decay-time resolution, decay-time and angular detection efficiencies, 
and also considers the contributions of the intermediate resonances to the $m(\pip\pim)$ spectrum.  
The PDF is convolved with a Gaussian function whose 
width is taken as the per-candidate decay-time uncertainty, derived from the uncertainties of the measured momentum and decay distance of $\Bd$ mesons and calibrated using control samples. 
The average decay-time resolution for signal candidates is determined to be $41\,\text{fs}$, which is significantly smaller 
than the $\Bd$ oscillation period and has a negligible impact on the measurement.

The acceptance, the reconstruction and the selection generate a nonuniform efficiency as a function of the $\Bd$ decay time, decay angles, and the invariant $\pip\pim$ mass. 
In the following, the efficiency related to the decay angles and the invariant $\pip\pim$ mass is generically denoted as angular efficiency, and it is assumed to be independent of that of the decay time. 
A four-dimensional angular efficiency 
is parametrized by a combination of Legendre and spherical harmonic moments~\cite{LHCb-PAPER-2019-003}, 
using simulated signal candidates kinematically corrected to match data.

The decay-time efficiency is determined with a well-established data-driven method~\cite{LHCb-PAPER-2019-003,LHCb-PAPER-2019-013}, using the control channel $\decay{\Bd}{\jpsi\Kstarz}$, which is topologically similar to the signal decay. The decay-time efficiency is modeled by a cubic spline function, determined from the decay-time 
distribution of selected candidates divided by that of generated candidates, 
where the latter decay-time distribution is modeled by an exponential distribution with the \Bd lifetime fixed to its known value~\cite{PDG2024}, 
convolved with a Gaussian resolution function whose parameters are determined from the resolution study described above. 
The kinematic differences between signal and control modes are corrected using the corresponding simulated samples.

The flavor of the \Bd meson at production is identified with two independent flavor-tagging algorithms, 
the opposite-side (OS) tagger~\cite{LHCb-PAPER-2011-027} and same-side (SS) tagger~\cite{LHCb-PAPER-2016-039}. 
Both methods provide a tagging decision $Q$ and an imperfectly estimated mistag probability $\kappa$ for each \Bd meson. 
The decision $Q$ takes values of $+1$, $-1$, or $0$, corresponding to classifications as \Bd, \Bdb, or untagged, respectively.
The calibration of the mistag probability $\omega(\kappa)$ follows the strategy in Ref.~\cite{LHCb-PAPER-2019-003}, 
using a linear function for each algorithm. The OS calibration is performed using $\decay{\Bu}{\jpsi\Kp}$ decays, 
with $\omega$ determined from the number of correct and wrong tagging decisions. 
The SS calibration employs $\decay{\Bd}{\jpsi\Kstarz}$ decays, 
whose flavors at decay are determined by the charge of final states, 
and $\omega$ is estimated by fitting the decay-time distribution.
The SS tagging algorithm is trained on samples whose selection biases the decay time and decay angles~\cite{LHCb-PAPER-2016-039}, resulting in different 
detection efficiencies for tagged and untagged candidates. 
The effective tagging power, defined as $\epsilon_{\rm tag}(1-2\omega)^2$ with $\epsilon_{\rm tag}$ being the tagging efficiency, is measured to be $4.5\%$ for the combined OS and SS taggers in the full \mbox{Run~2} dataset.
To simplify the analysis, untagged candidates representing 12\% of the signal are excluded in the fit as they are insensitive to the \CP violation.

The $m(\pip\pim)$ spectrum is modeled with six resonances, using the same configuration of amplitude magnitudes and relative phases for each polarization component as in Ref.~\cite{LHCb-PAPER-2014-058}. The $f_0(500)$ resonance is modeled using a Bugg function, with its parameters fixed to the values of the solution (iii) proposed in Ref.~\cite{Bugg:2006gc}. 
The $f_0(1270)$ and $\omega(782)$ resonances are described by relativistic Breit--Wigner (RBW) functions~\cite{PDG2024}, while the $\rho(770)^0$, $\rho(1450)^0$ and $\rho(1700)^0$ resonances are described by Gounaris--Sakurai functions~\cite{Gounaris:1968mw}. 
Their pole masses and decay widths are constrained to the world-average values~\cite{PDG2024}. 
The $\rho(770)^0$ resonance has an independent set of \CP-violation parameters $2\betaeff$ and $\abs{\lambda}$, 
while all the other resonances share another common set. 
The mass difference, decay width and decay width difference of \Bd mass eigenstates,
as well as the \Bd production asymmetry are shared across all the resonances. 
Among these, the first two parameters are constrained to their 
known values~\cite{PDG2024}, while the latter two are fixed to zero. 
This study reports two fits, where the parameters $2\betaeff$ and $\abs{\lambda}$ are either identical or not across all polarization amplitudes ($A_i$, $i=0,\perp,\parallel$), which are 
defined in the transversity basis~\cite{LHCb-PAPER-2014-058,Zhang:2012zk}. The polarization-independent 
fit is used later to constrain $\Delta\phis$, while the polarization-dependent one accounts for potential differences in the hadronization dynamics among the polarization states. 
The decay-time resolution parameters, as well as the coefficients of the angular and decay-time efficiencies, 
are fixed in the fit, while the flavor-tagging parameters are constrained to the calibration results.  

\begin{table}[tb]
  \centering
  \caption{
    \small
    Physics parameters of interest determined from the polarization-independent and \mbox{-dependent} 
    fits~\cite{LHCb-PAPER-2014-058,Zhang:2012zk}, where the first uncertainty is statistical and the second 
    systematic. The statistical correlation coefficient between $2\beta^{\rm eff}_{c\bar{c}d}$ and $\abs{\lambda}$ is $0.203$, while those for the polarization-dependent results are given 
    in the End Matter. 
  }
  \label{tab_fitres}
  \begin{tabular}{l@{\hskip 10pt}l@{\hskip 10pt}l@{\hskip 10pt}l}
  \hline
  Parameter & \multicolumn{3}{c}{Value}\\
  \hline
             $2\betaeff$ (\rad) & $0.710$ & $\pm\,0.084$ & $\pm\,0.051$ \\
                $\abs{\lambda}$ & $1.019$ & $\pm\,0.034$ & $\pm\,0.024$ \\
  \hline
                   $2\betaeffz$ \,(\rad) & $0.706$ & $\pm\,0.086$ & $\pm\,0.070$\\
                   $2\betaeffl$ \,(\rad) & $0.674$ & $\pm\,0.097$ & $\pm\,0.053$\\
                   $2\betaeffe$ \,(\rad) & $0.696$ & $\pm\,0.096$ & $\pm\,0.066$\\
                     $\abs{\lambda_{0}}$ & $1.004$ & $\pm\,0.045$ & $\pm\,0.056$\\
             $\abs{\lambda_{\parallel}}$ & $0.996$ & $\pm\,0.072$ & $\pm\,0.068$\\
                 $\abs{\lambda_{\perp}}$ & $1.203$ & $\pm\,0.154$ & $\pm\,0.080$\\
  \hline
  \end{tabular}
  \end{table}

\begin{figure}[ptb]
  \centering
  \begin{center}
     \includegraphics[width=0.65\linewidth]{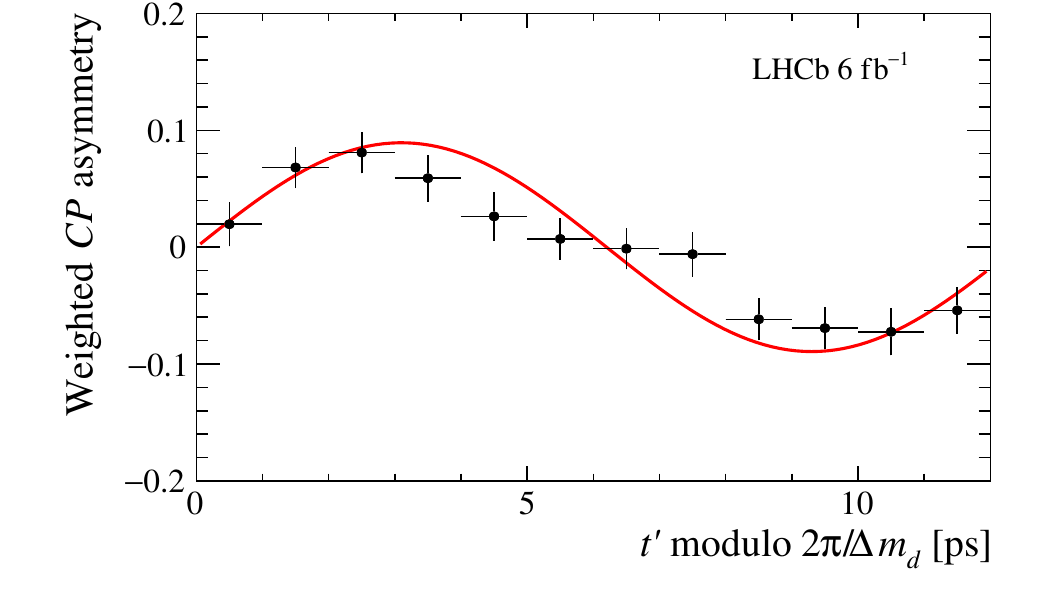}
     \vspace*{-0.5cm}
  \end{center}
  \caption{\small
    Weighted \CP asymmetry as a function of the transformed decay-time and phase-space variable $t^{\prime}$ over one oscillation period in the background-subtracted \mbox{$B^{0}\xspace\!\to{J\mskip-3mu/\mskip-2mu\psi}\pi^{+}\pi^{-}$} decays, together with the fit projection. 
  }
     \label{fig_bd_fittime}
\end{figure}

\begin{figure*}[ptb]
  \centering
     \includegraphics[width=0.45\linewidth]{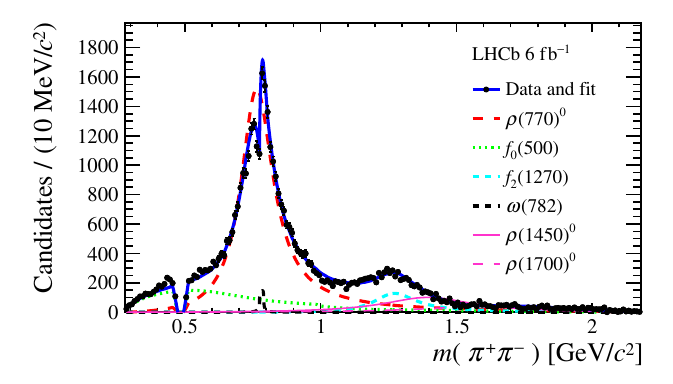}
     \includegraphics[width=0.45\linewidth]{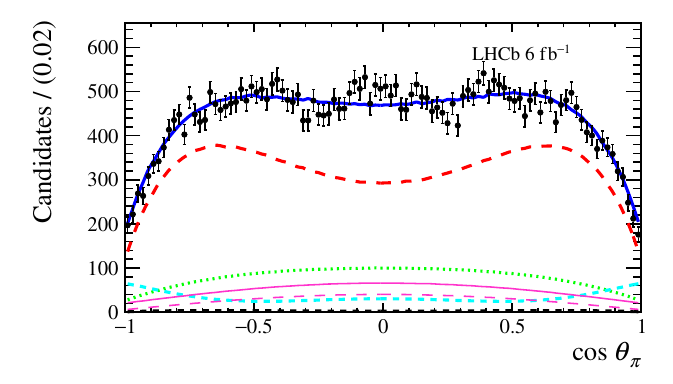} \\
     \includegraphics[width=0.45\linewidth]{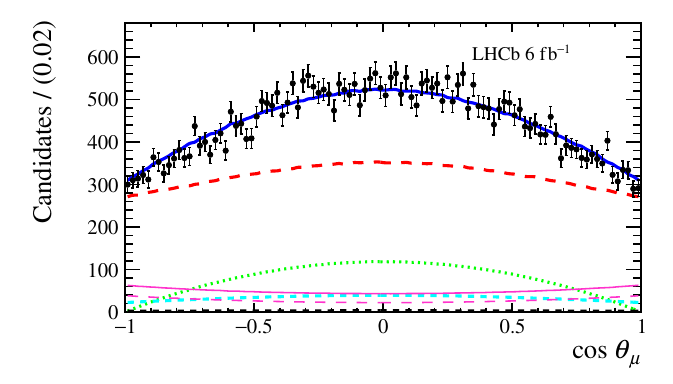}
     \includegraphics[width=0.45\linewidth]{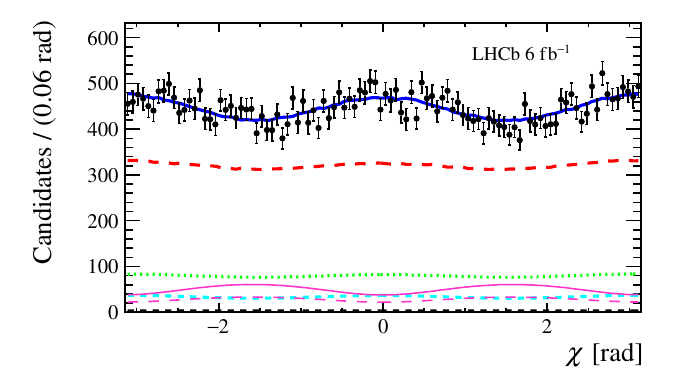}
     \vspace*{-0.5cm}
  \caption{\small 
      Background-subtracted distributions of $m(\pi^{+}\pi^{-})$ and three angular variables of the 
      \mbox{$B^{0}\xspace\!\to{J\mskip-3mu/\mskip-2mu\psi}\pi^{+}\pi^{-}$} decays, together with the fit projections. 
  }
     \label{fig_bd_fitamp}
\end{figure*}

Potential difference in the angular efficiency between simulated samples and data, arising from mismatches in the momenta of final-state particles, is corrected using an iterative fit as detailed in Ref.~\cite{LHCb-PAPER-2019-013}.
The measured \CP-violation parameters are summarized in \tabref{tab_fitres}, 
where statistical uncertainties are estimated using the bootstrapping method~\cite{Efron:1979bxm,Langenbruch:2019nwe}. 
This approach allows the uncertainties from the $\jpsi\pip\pim$ mass fit, including the uncertainties on the contribution of the partially reconstructed background, to be propagated into the statistical uncertainties of the \CP-violation parameters. 
All results are mutually consistent, showing no evidence for a polarization-dependent effect. 
The statistical significance of a nonzero value of 2\betaeff is found to be around ten Gaussian standard deviations using Wilks's theorem~\cite{Wilks:1938dza},  
establishing the first observation of \CP violation in the $\decay{\Bd}{\jpsi\rho(770)^0}$ decays. 
The corresponding time-dependent \CP asymmetry varies at different points in the phase space, leading to a dilution on this asymmetry. 
To reduce this dilution, a variable $t^{\prime}$, determined as the decay time adjusted by a factor dependent on phase-space position, is used. 
As described in the {\hyperlink{sec_endmatter}{\color{black} End Matter}}, it aligns 
the oscillation phases at different phase-space positions to reduce the dilution on 
the \CP asymmetry. The resulting \CP asymmetry is shown in \figref{fig_bd_fittime}. 
The background-subtracted distributions of $m(\pip\pim)$ and the decay angles, along 
with their fit projections, are shown in \figref{fig_bd_fitamp}. 
As a cross-check, the fit is repeated on subsets of the data, divided by the year of data-taking, magnet polarity, trigger and tagging category. 
In all cases, the obtained results are consistent with the baseline results.


The total systematic uncertainties reported in \tabref{tab_fitres} are obtained by summing the 
individual contributions in quadrature. These contributions are described below and 
summarized in the {\hyperlink{sec_endmatter}{\color{black} End Matter}}.
The uncertainties of input parameters constrained in the fit are already incorporated into the 
statistical uncertainties of the fit results. 
The impact of the decay-time resolution is found to be 
negligible on the fit results. 
The systematic uncertainties from the decay width difference and production asymmetry of the \Bd meson are evaluated by varying their values by one 
standard deviation, taken as the quadrature sum of the central values and uncertainties reported by Refs.~\cite{PDG2024} and~\cite{LHCb-PAPER-2023-013}, respectively.

Systematic uncertainties related to the $\jpsi\pip\pim$ mass fit model are evaluated by two 
alternative fits using a double-sided Crystal Ball function~\cite{Skwarnicki:1986xj} for 
the signal and a fourth-order polynomial function for the combinatorial background, respectively. 
Potential differences between the sideband candidates used for background subtraction and the true background under the \Bd peak are conservatively accounted for by retaining $20\%$ of the combinatorial background candidates in the fit. 
Systematic uncertainties related to the resonance model are evaluated using RBW functions for the $\rho$ resonances and the Bugg function, with solution (ii) from Ref.~\cite{Bugg:2006gc}, for the $f_0(500)$ resonance. 
The impact of the configuration of \CP-violation parameters is assessed by allowing independent parameters for each resonance in the fit. Alternative fits are performed including either an $f_0(980)$, $f_0(1500)$, $f_0(1700)$, or a nonresonant component, and the maximum observed variations are assigned as systematic uncertainties.

Systematic uncertainties from the correction of the simulated samples in the determination of the angular efficiency are evaluated by changing the variables used to derive the corrections. 
Conservative systematic uncertainties from the iterative fit procedure are evaluated by comparing the results with and without the iterative correction.
Untagged candidates, which exhibit different detection efficiencies compared to tagged candidates, are excluded in the baseline fit. However, as tagging information is unavailable in the control samples, untagged candidates are retained in the acceptance studies. An alternative fit that includes the untagged candidates is performed to assess the impact of this effect.

Systematic uncertainties due to the limited sizes of the control samples for the angular and decay-time efficiencies are estimated by varying the corresponding coefficients according to their statistical covariance matrices, where the decay-time contribution is found to be negligible. 
Varying the parametrization models of these efficiencies yields negligible changes. The impact of neglecting correlations between the angular and decay-time efficiencies is assessed using pseudoexperiments.

The results in \tabref{tab_fitres} and Ref.~\cite{LHCb-PAPER-2014-058} are consistent. Their combination, performed accounting for the correlations of systematic uncertainties, yields $2\betaeff= 0.718 \pm 0.088\rad$ and 
$\abs{\lambda}= 1.033 \pm 0.035$, with a correlation coefficient of $0.11$. 
A summary of the LHCb measurements of these parameters is shown in \figref{fig_combplot} 
in the {\hyperlink{sec_endmatter}{\color{black} End Matter}}. 
Parametrizing the magnitude and phase of the strong parts of the effective penguin amplitude relative to the tree amplitude as $ae^{i\theta}$ in the $\decay{\Bd}{\jpsi\rho(770)^0}$ and 
$a^{\prime}e^{i\theta^{\prime}}$ in the $\decay{\Bs}{\jpsi\phi(1020)}$ decays, and assuming these two ratios are identically quantified, 
the penguin shift $\Delta\phis$ can be estimated through~\cite{LHCb-PAPER-2014-058} 
\begin{equation*}
  \begin{aligned}
  \label{eqn_peng_phis2}
  \Delta\phis & = -{\rm arg}\mbracket{
  \frac{(\lamp{e}^{2i\gamma}-1)+\epsilon(\lamp-1)}
  {(\lamp{e}^{2i\gamma}-1)+\epsilon(\lamp-1){e}^{2i\gamma}}
  },
\end{aligned}
\end{equation*}
with
\begin{equation*}
  \begin{aligned}
  \lamp & \equiv \abs{\lambda}e^{-i(2\betaeff-2\betaccs)} 
        = \frac{1-ae^{i\theta}e^{-i\gamma}}{1-ae^{i\theta}e^{i\gamma}}, 
  \end{aligned}      
\end{equation*}
where $\epsilon=\nu^2/(1-\nu^2)$ is the Cabibbo-suppression factor~\cite{Barel:2020jvf}. 
The Wolfenstein parameter $\nu=0.22501 \pm 0.00068$ 
and the Cabibbo–Kobayashi–Maskawa (CKM) angle $\gamma=(65.7\pm3.0)\degrees$ are taken from Ref.~\cite{PDG2024}, 
and the \CP-violating phase $2\betaccs$ from measurements of $\decay{\Bd}{\jpsi\KS}$ decays~\cite{LHCb-PAPER-2023-013}.
The penguin contribution $\Delta\phis$ is determined to be $\ValDphis\mrad$, 
with an uncertainty significantly smaller than the current precision on $\phis$. 
A study of SU(3) flavor symmetry breaking is performed by simultaneously scanning the amplitude magnitude ratio $a/a^{\prime}$
in the range $[0.5, 1.5]$ and the phase difference \mbox{$\theta\!-\!\theta^{\prime}$} over $[-180, 180]^{\circ}$. 
As shown in the Supplemental Material~\cite{Supplemental_material}, the resulting $\Delta\phis$ values vary within 
the range $[-7.8,7.8]\mrad$ and its corresponding uncertainty increases to at most $6.9\mrad$, 
indicating that the uncertainty induced by the breaking of SU(3) flavor symmetry may exceed the precision of the estimate itself.

In summary, time-dependent \CP violation in $\decay{\Bd}{\jpsi\rho(770)^0}$ decays is observed for the first time. 
From a time-dependent amplitude analysis of $\decay{\Bd}{\jpsi\pip\pim}$ decays using the full Run~2 dataset collected by the \lhcb experiment, the \CP-violation parameters in the $\decay{\Bd}{\jpsi\rho(770)^0}$ process are measured to be
\mbox{$2\betaeff=0.710 \pm 0.084 \pm 0.051 \rad$} and \mbox{$\abs{\lambda}=1.019 \pm 0.034 \pm 0.024 $}. 
These results are consistent with, and two times more precise than the previous 
measurement~\cite{LHCb-PAPER-2014-058}.
Using SU(3) flavor symmetry to relate the sizes of the penguin contributions in $\decay{\Bd}{\jpsi\rho(770)^0}$ and $\decay{\Bs}{\jpsi\phi(1020)}$ decays, and combining the previous 
results in Ref.~\cite{LHCb-PAPER-2014-058}, the most stringent constraint on the penguin contribution is determined to be $\Delta\phis = \ValDphis\mrad$. 
The measured \CP-violation parameters will ultimately provide essential inputs to a global study of $B \to \jpsi X$ decays that aims to 
simultaneously determine the phases $2\beta$ and $\phis$ in the presence of 
penguin pollution~\cite{Barel:2020jvf,DeBruyn:2025rhk}.



\section*{Acknowledgements}
%
%
\noindent We express our gratitude to our colleagues in the CERN
accelerator departments for the excellent performance of the LHC. We
thank the technical and administrative staff at the LHCb
institutes.
We acknowledge support from CERN and from the national agencies:
ARC (Australia);
CAPES, CNPq, FAPERJ and FINEP (Brazil); 
MOST and NSFC (China); 
CNRS/IN2P3 (France); 
BMFTR, DFG and MPG (Germany);
INFN (Italy); 
NWO (Netherlands); 
MNiSW and NCN (Poland); 
MCID/IFA (Romania); 
MICIU and AEI (Spain);
SNSF and SER (Switzerland); 
NASU (Ukraine); 
STFC (United Kingdom); 
DOE NP and NSF (USA).
We acknowledge the computing resources that are provided by ARDC (Australia), 
CBPF (Brazil),
CERN, 
IHEP and LZU (China),
IN2P3 (France), 
KIT and DESY (Germany), 
INFN (Italy), 
SURF (Netherlands),
Polish WLCG (Poland),
IFIN-HH (Romania), 
PIC (Spain), CSCS (Switzerland), 
and GridPP (United Kingdom).
We are indebted to the communities behind the multiple open-source
software packages on which we depend.
Individual groups or members have received support from
Key Research Program of Frontier Sciences of CAS, CAS PIFI, CAS CCEPP, 
Minciencias (Colombia);
EPLANET, Marie Sk\l{}odowska-Curie Actions, ERC and NextGenerationEU (European Union);
A*MIDEX, ANR, IPhU and Labex P2IO, and R\'{e}gion Auvergne-Rh\^{o}ne-Alpes (France);
Alexander-von-Humboldt Foundation (Germany);
ICSC (Italy); 
Severo Ochoa and Mar\'ia de Maeztu Units of Excellence, GVA, XuntaGal, GENCAT, InTalent-Inditex and Prog.~Atracci\'on Talento CM (Spain);
SRC (Sweden);
the Leverhulme Trust, the Royal Society and UKRI (United Kingdom).

Data availability—The data that support the findings of this article are openly available~\cite{LHCb-PAPER-2025-059-cds}.


\newcommand{\tabincell}[2]{\begin{tabular}{@{}#1@{}}#2\end{tabular}}
\clearpage
\appendix

\section*{End Matter}
\hypertarget{sec_endmatter}{}

The \CP asymmetry shown in \figref{fig_bd_fittime} is obtained by the following procedure. 
The time-dependent \CP asymmetry in the $\decay{\Bd}{\jpsi\pip\pim}$ decays is described by 
\begin{equation*}
\label{eqn_tdacp}
   \frac{\overline{\Gamma}(t)-\Gamma(t)}{\overline{\Gamma}(t)+\Gamma(t)}
   = S\sin(\dmd\,{t})-C\cos(\dmd\,{t}),
\end{equation*}
where $\dmd$ is the mass difference of $\Bd$ mass eigenstates~\cite{PDG2024}, 
and $S$ and $C$ are \CP-violation coefficients depending on $(\mpipi,\cosmu,\cospi,\chi)$~\cite{Zhang:2012zk}.
Since the asymmetry depends on the location in the $(\mpipi,\cosmu,\cospi,\chi)$ phase space, the 
overall asymmetry integrated over the phase space is diluted as $S$ or $C$ changes sign. 
Moreover, it is also diluted by the flavor tagging effects. 
To view the \CP asymmetry, the decay time $t$ is transformed to $t^{\prime} = t + \deriv t(\mpipi,\cosmu,\cospi,\chi)$ using 
\begin{equation*}
\label{eqn_tdacp_relation}
\begin{aligned}
	\cos(\dmd\,\deriv t) & = S/\sqrt{S^2+C^2},\quad 
    \sin(\dmd\,\deriv t) & = -C/\sqrt{S^2+C^2}, \\
\end{aligned}
\end{equation*}
where $S$ and $C$ are determined by the fit to data. 
The asymmetry is transformed to a single sine function with positive coefficient
\begin{equation*}
\label{eqn_tdacp_trans}
   \frac{\overline{\Gamma}(t^{\prime})-\Gamma(t^{\prime})}{\overline{\Gamma}(t^{\prime})+\Gamma(t^{\prime})}
   = \sqrt{\mean{S}^2+\mean{C}^2}\sin(\dmd\,{t^{\prime}}),
\end{equation*}
where $\mean{S}$ $(\mean{C})$ is the average values of $S$ $(C)$ integrating over the phase space. 
The new asymmetry is not diluted because only positive coefficients are summed, and 
the dilution due to flavor tagging effects still remain. 
The quantity $\dmd\,{t^{\prime}}$ can be in any $2\pi$ period, so $\dmd\,{t^{\prime}}$ is taken to be modulo of $2\pi$. 

\begin{figure}[!b]
    \centering
    \begin{center}
       \includegraphics[width=0.45\linewidth]{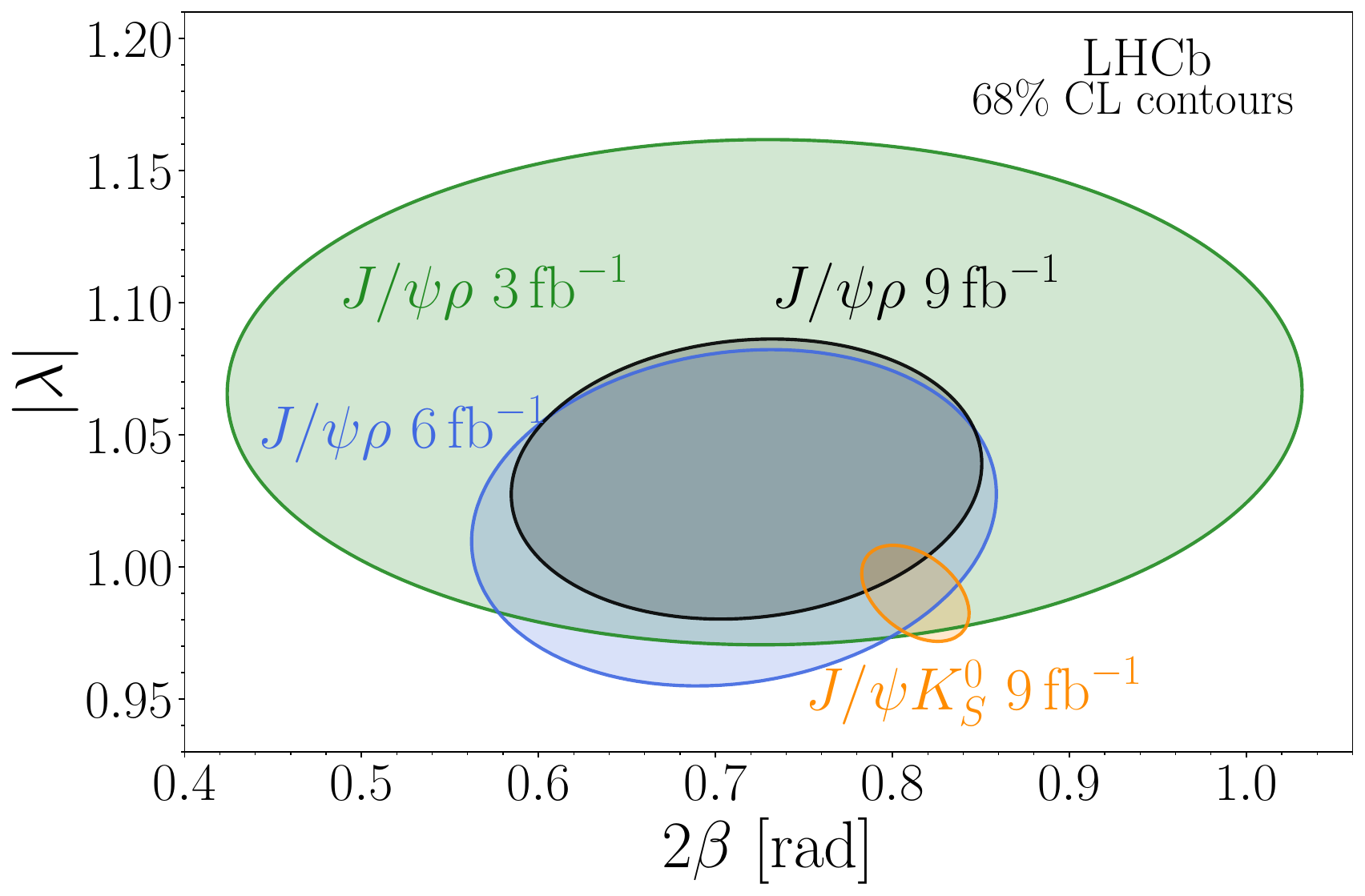}
       \vspace*{-0.5cm}
    \end{center}
    \caption{\small
      Comparison of LHCb measurements of the \CP-violation parameters in the 
      \mbox{$B^{0}\xspace\!\to{J\mskip-3mu/\mskip-2mu\psi}\rho(770)^0$} and 
      \mbox{$B^{0}\xspace\!\to{J\mskip-3mu/\mskip-2mu\psi}{K\xspace}^0_{\mathrm{S}}$} decays, 
      where $2\beta$ represents $2\beta^{\rm eff}_{c\bar{c}d}$ or $2\beta^{\rm eff}_{c\bar{c}s}$. 
    }
   \label{fig_combplot}
\end{figure}

The systematic uncertainties from different sources for the measured \CP-violation parameters are summarized in Tables~\ref{tab_syst} and \ref{tab_syst_polar}. 
The correlations for the polarization-dependent results are shown in Table~\ref{tab_correlation_polcpv}. 
Figure~\ref{fig_combplot} summarizes the \lhcb measurements of the \CP-violation parameters in the $\decay{\Bd}{\jpsi\rho(770)^0}$ and $\decay{\Bd}{\jpsi\KS}$ decays.

\begin{table}[!b]
    \begin{center}
      \caption{\small
        Systematic uncertainties of polarization-independent results from different sources.
      }
      \label{tab_syst}
      \begin{tabular}{lcc}
      \hline
      Source
      & \tabincell{c}{ $2\beta^{\rm eff}_{c\bar{c}d}$ \\ $(\aunit{mrad}\xspace\,)$ }
      & \tabincell{c}{ $\abs{\lambda}$ \\ $(\,10^{-3}\,)$ } \\
      \hline
        Decay width difference & \phantom{0}8.9 & \phantom{0}0.3 \\
        Production asymmetry & \phantom{0}1.7 & \phantom{0}1.4 \\
        $m(\jpsi\pip\pim)$ fit model & \phantom{0}0.1 & \phantom{0}0.0 \\
        Background subtraction & 11.3 & \phantom{0}1.5 \\
        Configuration of $\CP{V}$ parameters  & \phantom{0}6.5 & \phantom{0}1.3 \\
        Resonance model  & 13.8 & 12.5 \\
        Resonance contribution  & 38.9 & 18.5  \\
        Variables for kinematic correction & \phantom{0}6.8 & \phantom{0}1.2 \\
        Iterative fit & 17.0 & \phantom{0}2.0 \\
        Acceptance of untagged events & 11.0 & \phantom{0}8.3 \\
        Angular efficiency & \phantom{0}1.2 & \phantom{0}0.4 \\
        Efficiency correlation & 11.6 & \phantom{0}1.9 \\
        \hline
        \textbf{Total} & 50.5 & 24.1 \\
      \hline
      \end{tabular}
    \end{center}
\end{table}

\begin{table}[tb]\footnotesize
\begin{center}
  \caption{\small
    Systematic uncertainties of polarization-dependent results from different sources. 
  }
  \label{tab_syst_polar}
  \begin{tabular}{lcccccc}
  \hline
  Source 
  & \tabincell{c}{ $2\beta^{\rm eff}_{0}$          \\ $(\aunit{mrad}\xspace\,)$ }     
  & \tabincell{c}{ $2\beta^{\rm eff}_{\parallel}$  \\ $(\aunit{mrad}\xspace\,)$ }     
  & \tabincell{c}{ $2\beta^{\rm eff}_{\perp}$      \\ $(\aunit{mrad}\xspace\,)$ }     
  & \tabincell{c}{ $\abs{\lambda_{0}}$ \\ $(\,10^{-3}\,)$ }  
  & \tabincell{c}{ $\abs{\lambda_{\parallel}}$ \\ $(\,10^{-3}\,)$ }
  & \tabincell{c}{ $\abs{\lambda_{\perp}}$ \\ $(\,10^{-3}\,)$ } \\
  \hline

Decay width difference	    & \phantom{0}8.2 & \phantom{0}7.9 & 17.7 & \phantom{0}0.7 & \phantom{0}2.8 & \phantom{0}6.7 \\
Production asymmetry	    & \phantom{0}2.2 & \phantom{0}1.8 & \phantom{0}2.8 & \phantom{0}2.4 & \phantom{0}0.7 & \phantom{0}3.5 \\
$m(\jpsi\pip\pim)$ fit model	    & \phantom{0}0.2 & \phantom{0}0.2 & \phantom{0}0.5 & \phantom{0}0.1 & \phantom{0}0.6 & \phantom{0}0.2 \\
Background subtraction	    & \phantom{0}9.5 & 12.5 & 14.0 & \phantom{0}1.3 & 11.6 & \phantom{0}1.5 \\
Configuration of $\CP{V}$ parameters 	    & 37.6 & 27.3 & 35.4 & 42.1 & 65.3 & 32.2 \\
Resonance model 	    & 18.4 & 14.8 & 14.1 & 24.8 & 14.5 & \phantom{0}7.5 \\
Resonance contribution 	    & 43.9 & 32.8 & 43.7 & 24.2 & \phantom{0}0.5 & 58.1 \\
Variables for kinematic correction	    & \phantom{0}7.6 & \phantom{0}7.3 & \phantom{0}6.5 & \phantom{0}1.9 & \phantom{0}1.3 & \phantom{0}5.5 \\
Iterative fit	    & 18.4 & 17.9 & 17.4 & \phantom{0}1.6 & \phantom{0}1.9 & 15.2 \\
Acceptance of untagged events	    & 20.6 & \phantom{0}7.0 & \phantom{0}3.1 & 11.1 & \phantom{0}3.0 & 40.1 \\
Angular efficiency	    & \phantom{0}1.5 & \phantom{0}1.1 & \phantom{0}1.4 & \phantom{0}0.8 & \phantom{0}0.4 & \phantom{0}1.5 \\
Efficiency correlation	    & 14.3 & 11.1 & 13.5 & \phantom{0}3.5 & \phantom{0}1.8 & \phantom{0}6.6 \\
\hline
\textbf{Total} & 69.8 & 53.0 & 66.4 & 55.9 & 68.0 & 80.3 \\
  \hline
  \end{tabular}
\end{center}
\end{table}

\begin{table}[tb]
   \begin{center}
	\caption{\small
	   \label{tab_correlation_polcpv}
	   Statistical correlations between the polarization-dependent \CP-violation parameters. 
	}
	\begin{tabular}{ccccccc}
	   \hline
      Parameter &$\quad 2\beta^{\rm eff}_{0}$ &$\quad 2\beta^{\rm eff}_{\parallel}$ &$\quad 2\beta^{\rm eff}_{\perp}$ &$\quad \abs{\lambda_{0}}$ &$\quad \abs{\lambda_{\parallel}}$
 &$\quad \abs{\lambda_{\perp}}$\\
      \hline
            $\quad 2\beta^{\rm eff}_{0}$ &$\phantom{-}1.000$ &$\phantom{-}0.784$ &$\phantom{-}0.767$ &$\phantom{-}0.311$ &$-0.021$ &$\phantom{-}0.027$\\
            $\quad 2\beta^{\rm eff}_{\parallel}$ &$\phantom{-}0.784$ &$\phantom{-}1.000$ &$\phantom{-}0.830$ &$\phantom{-}0.094$ &$\phantom{-}0.036$ &$-0.070$\\
            $\quad 2\beta^{\rm eff}_{\perp}$ &$\phantom{-}0.767$ &$\phantom{-}0.830$ &$\phantom{-}1.000$ &$\phantom{-}0.095$ &$\phantom{-}0.108$ &$-0.097$\\
     $\quad \abs{\lambda_{0}}$ &$\phantom{-}0.311$ &$\phantom{-}0.094$ &$\phantom{-}0.095$ &$\phantom{-}1.000$ &$-0.074$ &$\phantom{-}0.105$\\
$\quad \abs{\lambda_{\parallel}}$ &$-0.021$ &$\phantom{-}0.036$ &$\phantom{-}0.108$ &$-0.074$ &$\phantom{-}1.000$ &$-0.388$\\
 $\quad \abs{\lambda_{\perp}}$ &$\phantom{-}0.027$ &$-0.070$ &$-0.097$ &$\phantom{-}0.105$ &$-0.388$ &$\phantom{-}1.000$\\
      \hline
	\end{tabular}
   \end{center}
\end{table}
 
\clearpage
\addcontentsline{toc}{section}{References}
\setboolean{inbibliography}{true}
\bibliographystyle{LHCb}
\bibliography{main,standard,LHCb-PAPER,LHCb-CONF,LHCb-DP,LHCb-TDR}


\clearpage
\centerline
{\large\bf LHCb collaboration}
\begin
{flushleft}
\small
R.~Aaij$^{38}$\lhcborcid{0000-0003-0533-1952},
A.S.W.~Abdelmotteleb$^{58}$\lhcborcid{0000-0001-7905-0542},
C.~Abellan~Beteta$^{52}$\lhcborcid{0009-0009-0869-6798},
F.~Abudin{\'e}n$^{60}$\lhcborcid{0000-0002-6737-3528},
T.~Ackernley$^{62}$\lhcborcid{0000-0002-5951-3498},
A. A. ~Adefisoye$^{70}$\lhcborcid{0000-0003-2448-1550},
B.~Adeva$^{48}$\lhcborcid{0000-0001-9756-3712},
M.~Adinolfi$^{56}$\lhcborcid{0000-0002-1326-1264},
P.~Adlarson$^{86}$\lhcborcid{0000-0001-6280-3851},
C.~Agapopoulou$^{14}$\lhcborcid{0000-0002-2368-0147}
% ...

{\footnotesize \it

% ...
$^{14}$Universit{\'e} Paris-Saclay, CNRS/IN2P3, IJCLab, Orsay, France\\
% ...
$^{38}$Nikhef National Institute for Subatomic Physics, Amsterdam, Netherlands\\
% ...
$^{48}$Instituto Galego de F{\'\i}sica de Altas Enerx{\'\i}as (IGFAE), Universidade de Santiago de Compostela, Santiago de Compostela, Spain\\
% ...
$^{52}$Physik-Institut, Universit{\"a}t Z{\"u}rich, Z{\"u}rich, Switzerland\\
% ...
$^{56}$H.H. Wills Physics Laboratory, University of Bristol, Bristol, United Kingdom\\
% ...
$^{58}$Department of Physics, University of Warwick, Coventry, United Kingdom\\
% ...
$^{60}$School of Physics and Astronomy, University of Edinburgh, Edinburgh, United Kingdom\\
$^{61}$School of Physics and Astronomy, University of Glasgow, Glasgow, United Kingdom\\
$^{62}$Oliver Lodge Laboratory, University of Liverpool, Liverpool, United Kingdom\\
% ...
$^{70}$Syracuse University, Syracuse, NY, United States\\
% ...
$^{86}$Department of Physics and Astronomy, Uppsala University, Uppsala, Sweden, associated to $^{61}$\\
% ...
\medskip
}
\end{flushleft}

\end{document}